\documentclass[10pt]{article}
\usepackage{amsmath,amssymb}
\usepackage[pdftex]{graphicx}
\usepackage[hidelinks]{hyperref}
\usepackage{amsthm}
\usepackage{bm}
\usepackage{extarrows}
\usepackage{txfonts}
\usepackage{here}
\usepackage{authblk} 

\theoremstyle{definition}
 \newtheorem{dfn}{Definition}[section]
 \newtheorem{pro}[dfn]{Proposition}
 \newtheorem{lem}[dfn]{Lemma}
 \newtheorem{thm}[dfn]{Theorem}
 \newtheorem{exm}[dfn]{Example}
 \newtheorem{cor}[dfn]{Corollary}

\usepackage{mathrsfs}
\newcommand{\R}{\mathbb{R}}
\newcommand{\Z}{\mathbb{Z}}
\newcommand{\N}{\mathbb{N}}
\newcommand{\Bs}{\bm{s}}
\newcommand{\mfS}{\mathfrak{S}}
\newcommand{\mcT}{\mathcal{T}}
\newcommand{\mcU}{\mathcal{U}}
\newcommand{\msA}{\mathscr{A}}
\newcommand{\rmb}{\mathrm{b}}
\newcommand{\rmc}{\mathrm{c}}

\date{}

\begin{document}
 \title{Solitons in 3-State Mealy Automata}
 \author[1]{Atsushi Maeno\thanks{maeno.atsushi.45s@kyoto-u.ac.jp}}
 \author[1]{Satoshi Tsujimoto\thanks{tsujimoto.satoshi.5s@kyoto-u.jp}}
 \author[2]{Fumitaka Yura\thanks{f-yura@musashino-u.ac.jp}}
 \affil[1]{Graduate School of Informatics, Kyoto University}
 \affil[2]{Faculty of Engineering, Musashino University}
 \vspace{10pt}
 \maketitle
 
 \begin{abstract}
  Box--ball systems (BBS) are integrable systems with soliton solutions and other good properties. We will search for automata that belong to the same class as BBS automata by introducing some classes of automata through the features of BBS automaton. In particular, we would like to classify 3-state automata over a 2-letter alphabet. 
 \end{abstract}
 \noindent{\it Keywords\/}: Soliton, Mealy automata, box--ball system

 \section{Introduction}
\begin{figure}[ht] \centering
  \includegraphics[width=0.5\linewidth]{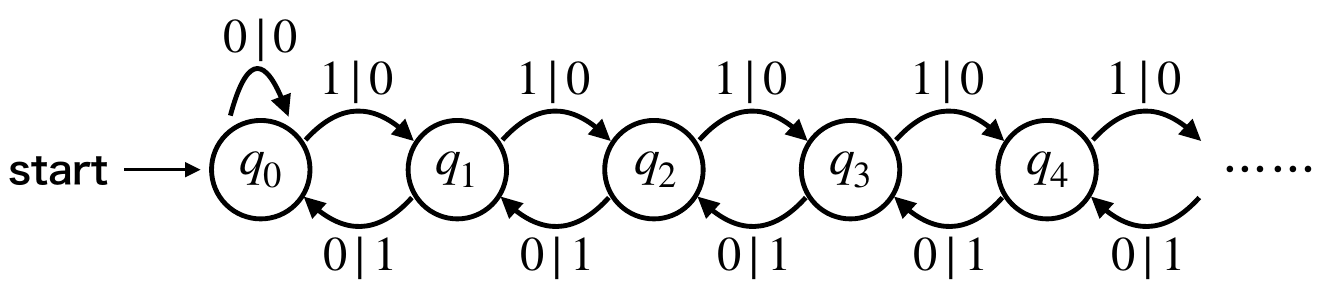}
  \caption{The Mealy automaton representing Takahashi--Satsuma's BBS. `start' means that $q_0$ is the initial state. }
  \label{Fig: BBS_Mealy}
 \end{figure}

 Recently we noticed that Takahashi--Satsuma's box--ball system (BBS)~\cite{TS90} can be represented as a Mealy automaton with infinitely many states~\cite{TKZ14} (shown in Figure\ref{Fig: BBS_Mealy}). It is worth exploring the interrelation between the BBS and Mealy automata. In this study, we examine Mealy automata from the perspective of the BBS. The Mealy automaton that describes the time evolution of the BBS exhibits several notable properties, including being a conserved system (particle-preserving), bijective, transitive, and locally interacting. In this paper, we first search for the soliton Mealy automata of an alphabet size of two and state set size of one, two, and three based on certain {\it key properties} of the BBS automaton using a computer. Subsequently, we discuss the linearizability of these automata. 

 \section{Mealy Automata}
 In this section, we recall the definition of Mealy automata. Let $Q$ and $S$ be a non-empty set of states and a non-empty set of letters called {\it alphabet}, respectively. 
 \begin{dfn}
  We introduce a transition function $\varphi$ and an exit function $\psi$: 
  \begin{eqnarray}
   \varphi: Q\times S\to Q; (q, s)\mapsto\varphi(q, s),\\
   \psi: Q\times S\to S; (q, s)\mapsto\psi(q, s).
  \end{eqnarray}
  Then, we also introduce the mappings $\varphi_s: Q\to Q; q\mapsto\varphi(q, s)$ for $s\in S$ and the mapping $\psi_q: S\to S; s\mapsto\psi(q, s)$ for $q\in Q$. 
 \end{dfn}

 \begin{dfn}
  An automaton $\msA$ is defined by a quadruple $(Q, S, \varphi, \psi)$. The action of $\msA$ on the product of the set of states and the set of words with finite length $n$ is defined by
  \begin{equation}
   (\varphi, \psi):Q\times S^n\to Q\times S^n; (q; s_1s_2\cdots s_n)\mapsto (\tilde q; \tilde s_1 \tilde s_2 \cdots\tilde s_n), 
  \end{equation}
  where $q_1=q$, $q_{j+1}=\varphi(q_j, s_j), \tilde s_j = \psi(q_j, s_j) $ for $j=1,2,\dots,n$ and $\tilde q=q_{n+1}$. The automaton with initial state $q^\prime\in Q$ is denoted by
   $\msA_{q^\prime}=(Q, S, \varphi, \psi; q^\prime)$ which acts on $S^n$ as
   \begin{equation}
    \psi_{q^\prime} : S^n\to S^n ; s_1\dots s_n \mapsto\tilde{s}_1 \tilde{s}_2\dots\tilde{s}_n.
   \end{equation}
  \end{dfn}
  The automaton $(Q, S, \varphi, \psi)$ is called the automaton of the type ($|Q|, |S|$). For a given pair of $(Q, S)$, $\varphi$ and $\psi$ are uniquely determined by the pair of $|Q|\times|S|$ matrices whose entries are in $Q$ and $S$, respectively. Thus the total number of automata of the type $(|Q|, |S|)$ is given by $(|Q|\times|S|)^{|Q|\times|S|}$.

 \begin{dfn}
  Assume $Q=\{q_0, q_1, \ldots, q_{m-1}\}$ and $S=\{0, 1, \dots, n-1\}$. The automaton $\msA=(Q, S, \varphi, \psi)$ is determined by a quadruple of integers $[m, n, k, l]$ as follows:
  \begin{itemize}
   \item Let $\varphi(q_i,j)=q_{f(i,j)}$ and $\psi(q_i,j)=g(i,j)$ where $f(i,j)$ and $g(i,j)$ are the elements of two $m\times n$ matrices, which are given by the $m$-ary expansion of $k$ and the $n$-ary expansion of $l$, as
   \begin{equation}
    \left(f(i,j)\right)_{0\le i<m, 0\le j<n}, \left(g(i,j)\right)_{0\le i<m, 0\le j<n}\in\mathrm{Mat}(m,n),
   \end{equation}
   where 
   \begin{equation}
    k=\sum_{i=0}^{m-1}\sum_{j=0}^{n-1} f(i,j) m^{m\,n -i \,n -j-1}, \quad
    l=\sum_{i=0}^{m-1}\sum_{j=0}^{n-1} g(i,j) n^{m\,n -i \,n -j-1}.
   \end{equation}
  \end{itemize}
  We call this automaton MA-$[m, n, k, l]$. 
 \end{dfn}

 \begin{exm}
  $m=3, n=2$, $104=1 \cdot 3^4 + 2 \cdot 3^2+1 \cdot 3^1+2 \cdot 3^0$, $11=1 \cdot 2^3 + 1 \cdot 2^1+1 \cdot 2^0$. 
  \begin{equation}
   \mbox{MA-}[3, 2,  104, 11] \mapsto\left[Q=\{q_0,q_1,q_2\},S=\{0,1\},\left( \begin {array}{cc} 0&1\\ 0&2\\ 1&2\end {array} \right) , \left( \begin {array}{cc} 0&0\\ 1&0\\ 1&1 \end {array} \right) \right]
  \end{equation}
  This automaton is the BBS with a carrier capacity of two, described later. 
 \end{exm}

 Next, we introduce the isomorphism and minimality of Mealy automata. 
 \begin{dfn}
  If there exists a pair of permutations $(\sigma, \rho) \in \mfS_{|Q|} \times \mfS_{|S|}$ such that 
  \begin{eqnarray}
   \begin{cases}
    \tilde \varphi(q,s)=\sigma(\varphi(\sigma^{-1}(q),\rho^{-1}(s))),\\
   \tilde \psi(q,s)=\rho(\psi(\sigma^{-1}(q),\rho^{-1}(s))),\qquad\mbox{for all $q\in Q, s \in S$,}
   \end{cases}
  \end{eqnarray}
  then we say that the automaton $(Q, S, \tilde\varphi, \tilde\psi)$  is {\it isomorphic} to the automaton  $(Q, S, \varphi, \psi)$.
 \end{dfn}

 \begin{dfn}
  $\msA$  is {\it reduced} or {\it minimal} if for any distinct states $q\in Q$ and $r\in Q$, there exists $\Bs\in S^*$ such that $\psi(q, \Bs) \ne \psi(r, \Bs)$.
 \end{dfn}
 Here, $S^*$ is the {\it Kleene closure} of $S$, which is an infinite set containing the empty string $\varepsilon$ and all possible concatenations of letters in $S$. If $\msA$ is not minimal, the behavior of $\msA$ can be described by an automaton that has fewer states than $\msA$. 

 \section{\texorpdfstring{BBS-C($k$): The box--ball system with finite carrier capacity $k$}{}}
\begin{figure}[ht]\centering
  \includegraphics[width=0.5\linewidth]{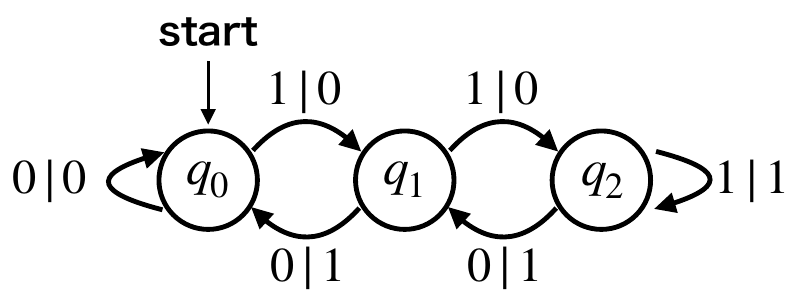}
  \caption{The MA-$[3, 2, 104, 11]$ of the BBS with a carrier capacity of $k=2$. }
  \label{Fig: BBS-C(2)}
 \end{figure}

 The BBS with a carrier capacity of $k$ (BBS-C($k$)), which is proposed by Takahashi and Matsukidaira~\cite{TM97}, can be described as a Mealy automaton (shown in Figure~\ref{Fig: BBS-C(2)}). The MA-[3,2,104,11] is equivalent to the BBS-C(2). Kuniba et al. proved that the time evolution of the BBS-C($k$) is linearized by the Kerov--Kirillov--Reshetikhin bijection~\cite{KOSTY06}. Additionally, Kakei et al. provided an alternative linearization method based on the 01-arc lines~\cite{KNTW18}. 
 
 \subsection{\texorpdfstring{Properties of the BBS-C($k$)}{}}
 Here we introduce several important properties of Mealy automata: transitive, time-reversible, particle-preserving, and locally interacting. 

 \begin{dfn}
  The transition function $\varphi: Q\times S\to Q$ is {\it transitive} if for every pair of two states $q$ and $q^\prime$ there exists a finite sequence $(s_1, s_2, \ldots, s_n)\in S^n$ such that $\varphi_{s_n}\circ \varphi_{s_{n-1}}\circ \cdots\circ\varphi_{s_1}(q)=q^\prime$. In this paper, we say the automaton $(Q, S, \varphi, \psi)$ is transitive if $\varphi$ is transitive. 
 \end{dfn}
 In other words, the transitive Mealy automaton is strongly connected as a directed graph. The following lemma and proposition are derived from the graph's strong connectivity. We use them to check whether the automaton is transitive or not. 

 \begin{lem}
  Let $\varphi$ be a transition function. If $\varphi$ is transitive, there doesn't exist $q^\prime\in Q$ such that
  \begin{equation}
   q^\prime \notin \{\varphi(q,s) \mid q \in Q\backslash\{q^\prime\}, s \in S\}.
  \end{equation}
  \begin{proof}
   If $q^\prime$ is not in $\{\varphi(q,s) \mid q \in Q\backslash\{q^\prime\}, s \in S\}$, it is not possible to reach $q^\prime$ from any other state $q$, and $\varphi$ is not transitive.
  \end{proof}
 \end{lem}

 \begin{pro}
  The transition function $\varphi$ is transitive if and only if for all non-empty proper subset $R\subsetneq Q$, $\{\varphi(q, s) \mid q\in R, s\in S\}$ is not a subset of $R$.
  \begin{proof}
  First, we assume $\varphi$ is transitive, and let $R$ be a non-empty proper subset of $Q$. If $\{\varphi(q, s) \mid q\in R, s\in S\}$ is a subset of $R$, then it is not possible to reach $q^\prime\in Q\backslash R$ from $q\in R$.

  Inversely, we assume that for all $R\subsetneq Q$, $\{\varphi(q, s)\mid q\in R, s\in S\}$ is not a subset of $R$. For each $q\in Q$, let $R_q$ be the set of all reachable states from $q$ as $R_q=\{\varphi_{s_n}\circ\varphi_{s_{n-1}}\circ\cdots\circ\varphi_{s_1}(q)\mid n\in\Z_{\ge0}, s_i\in S\quad(i=1, 2, \ldots, n)\}$. Because of $\{\varphi(q^\prime, s)\mid q^\prime \in R_q, s\in S\}=R_q$, $R_q$ is equal to $Q$. Therefore $\varphi$ is transitive.
  \end{proof}
 \end{pro}

 Next, we discuss the time-reversibility of automata. There are two types of time-reversibility, \textrm{i.e.} (left-)invertible automaton and right-invertible automaton. 

 \begin{dfn}
  The automaton $(Q,S,\varphi,\psi)$ is said to be {\it (left-)invertible} if the mapping $\psi_q: S\to S; s \mapsto \psi(q,s)$ is a bijection on $S$ for all $q\in Q$.
 \end{dfn}

 Note that left-invertible automata lead to the automata group~\cite{GZ01, H82}. In the BBS-C($k$), the exit function at state $q_0$ outputs 0 regardless of the input, indicating it is not left-invertible. 

 \begin{dfn}
  The automaton $(Q,S,\varphi,\psi)$ is said to be {\it right-invertible} or {\it bijective} if the mapping $(q,s) \mapsto  (\varphi(q,s), \psi(q,s))$ is a bijection on $Q\times S$.
 \end{dfn}
 The total number of bijective automata of the type ${(m,n)}$ is $(m\times n)!$. 

 \begin{dfn}
  The automaton $\msA$ is said to be {\it particle-preserving} if there exists a weight function $w : Q \cup S\to \R$ such that the function $w$ is not identically constant for $s\in S$ and 
  \begin{equation}
   w(q)+w(s)  = w(\tilde q) + w(\tilde s)\quad \mbox{for all $(q,s) \in Q\times S$},
  \end{equation}
  where $\tilde q = \varphi(q,s)$ and $\tilde s = \psi(q,s)$.
 \end{dfn}

 Because we can consider whether an automaton is particle-preserving, even if it is not minimal, the condition that the weight function $w$ is identically constant for $q\in Q$ is unnecessary. Next, we define the vacuum alphabet and the final state set. 

 \begin{dfn}
  Let $F$ and $V$ be a non-empty subset of $Q$ and $S$, respectively, such that, for all $q, q^\prime \in F, q^{\prime\prime} \in Q \backslash F$ and $\Bs \in V^*$,
  \begin{equation}
   \varphi(q, \Bs) \ne q^{\prime\prime}, \quad \varphi(q, \Bs^\prime) = q^\prime
  \end{equation}
  where $\Bs^\prime$ is some element of $V^*$. We call $V$ {\it the vacuum alphabet} and $F$ {\it the final state set} with respect to $V$. In the case that the final state set with respect to $V$ is unique, then we call $F$ {\it the vacuum state set} and the pair $(F, V)$ {\it the vacuum pair}.
 \end{dfn}

 \begin{dfn}
  Let $V$ be a subset of $S$. We call $\msA=(Q,S,\varphi,\psi)$ the {\it locally interacting automaton} with respect to  $V$ if for all $\Bs_1, \Bs_2\in S^*$, there exist $\Bs, \Bs^\prime \in V^*$ and $q \in Q$ such that
  \begin{enumerate}\renewcommand{\labelenumi}{(\alph{enumi}).}
   \item the length of $\Bs_1\Bs$ equals to the length of $\Bs_2\Bs^\prime$, and 
   \item for all $\Bs_3\in S^*$, 
   \begin{equation}
    \psi( {\varphi(q,\Bs_1\Bs)}, \Bs_3 ) = \psi({\varphi(q,\Bs_2\Bs^\prime)}, \Bs_3 ).
   \end{equation} 
  \end{enumerate}
  If $\msA$ is not the locally interacting automaton, then we call it the non-locally interacting automaton with respect to $V$.
 \end{dfn}

 In a locally interacting automaton with initial state $q$, if any strings $\Bs_1, \Bs_3\in S^*$ are separated by a sufficiently long string $\Bs\in V^*$, then $\Bs_1$ and $\Bs_3$ do not influence each other under the action of $\msA_q$. 

 \begin{pro}
  Let $\msA=(Q, S, \varphi, \psi)$ be a minimal,  transitive, and locally interacting automaton. Then, $\msA$ has a unique final state set with respect to $V$.
  \begin{proof}
   Suppose that $\msA$ is a minimal, transitive, and locally interacting automaton. Let $q_1=\varphi(q, \Bs_1)$, and $q_2=\varphi(q, \Bs_2)$. Since $\msA$ is transitive, $q_1$ and $q_2$ take arbitrary states in $Q$ for $\Bs_1$ and $\Bs_2$. Therefore, for all $q_1, q_2\in Q$ and $\Bs_3\in S^*$ there exist $\Bs, \Bs^\prime\in V^*$ such that
   \begin{equation}
    \psi(\varphi(q_1, \Bs), \Bs_3) = \psi(\varphi(q_2, \Bs^\prime), \Bs_3).
   \end{equation} 
   The minimal automaton does not allow two different states ${\varphi(q_1,\Bs)}$ and ${\varphi(q_2,\Bs^\prime)}$ satisfying the relation above, thus for all $q_1,q_2$ there exist $\Bs, \Bs^\prime\in V^*$ such that ${\varphi(q_1,\Bs )}={\varphi(q_2,\Bs^\prime)}$. By introducing $q_0 = {\varphi(q_2,\Bs^\prime)}$, it is shown that for all $q_1$ there exist $\Bs\in V^*$ such that ${\varphi(q_1,\Bs)}=q_0$. Hence the final state set is unique.
  \end{proof}
 \end{pro}

 \subsection{Soliton Automata}
  Suppose that the vacuum alphabet is $V=\{0\}$ here and hereafter. Since we consider the particle-preserving automata, we take the initial state $q=q_0$ and suppose $q_0$ is in the final state set $F$ with respect to $V$. It is also possible to consider cases where the boundary conditions increase the number of balls in the system, in which the initial state can be set to a different state from $q_0$. We define the solitary wave and the soliton automata. 

 \begin{dfn}[solitary wave $v$]
  A string $v$ of a finite length is called the {\it solitary wave}, if there exists a pair of $(m,l) \in \N^2$ such that ${(\psi_{q_0})}^m(v\,0^{\N})=0^{\,l} v \,0^{\N}$, where $0^{\N}$ is the semi-infinite sequence $000\cdots$ and $0^l$ is the sequence of length $l$ as $\underbrace{0\cdots0}_{l}$. This pair of non-negative integers $(m,l)$ determines the (average) speed $\mbox{sp}(v)=l/m$ and the fundamental period of $v$, denoted by $\mbox{period}(v)$, is given by the smallest value of $m$.  
 \end{dfn}

 \begin{dfn}[soliton automata ${\msA}$]
  Let $\msA$ be the bijective automaton. For all $\Bs \in S^*$, there is some $t\in \N$ such that, for some $n \in \N$ and solitary waves $v_1,v_2,\ldots,v_n$,
  \begin{eqnarray}
   & \msA^{t}_{q_0}( \Bs 0^{\N}) = 0^{k_1} v_1 0^{k_2} v_2 0^{k_3} \cdots 0^{k_n} v_n 0^{\N}, \\
   & \mbox{sp}(v_1) < \mbox{sp}(v_2) < \cdots <\mbox{sp}(v_n),
  \end{eqnarray}
  where $k_2,\ldots,k_n \ge|Q|-1$.
 \end{dfn}
 The condition $k_2, \ldots, k_n \ge|Q|-1$ means that the interaction between the solitary waves has already finished. 

 \section{Classification of the 3-state Mealy automata over a 2-letter alphabet}
 In this section, we will classify the class of 3-state Mealy automata over a 2-letter alphabet, that is the class of automata of type ($Q=\{q_0,q_1,q_2\}, S=\{0,1\}, *, *)$. Since the number of automata of type $(|Q|, |S|)$ is only $(|Q|\times|S|)^{|Q|\times|S|}$, we can use a computer to check whether they satisfy the properties of being bijective, transitive, particle-preserving, and locally interacting. Before going to this class, it is better to state on the simpler classes (1) $Q=\{q_0\}, S=\{0,1\}$ and (2) $Q=\{q_0,q_1\}, S=\{0,1\}$: 

 \subsection{\texorpdfstring{Case $Q=\{q_0\}, S=\{0,1\}$}{}}
 The total number of automata in this class is just $(1 \times 2)^{1 \times 2}=4$: [1,2,0,0], [1,2,0,1], [1,2,0,2], [1,2,0,3]. Here the output binary sequences of MA-[1,2,0,0] (MA-[1,2,0,3]) are all zeros (all ones) for any binary input sequence. The MA-[1,2,0,2] is the permutation of 0s and 1s, thus the time-evolution behavior is blinking manner $0\to 1 \to 0 $ or $1\to 0 \to 1 $. The MA-[1,2,0,1] identically acts on the binary sequence. Hence the MA-[1,2,0,1] is the only Mealy automaton of type $(1,2)$ of bijective, transitive, and particle-preserving.

 \subsection{\texorpdfstring{Case $Q=\{q_0,q_1\}, S=\{0,1\}$}{}}
\begin{figure}[ht]\centering
  \includegraphics[width=0.37\linewidth]{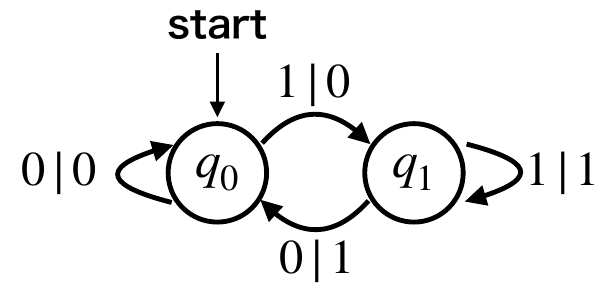}
  \caption{The bijective, transitive, particle-preserving, and minimal Mealy automaton of type (2, 2): MA-[2, 2, 5, 3] (BBS-C(1)). }
  \label{Fig: BBS-C(1)}
 \end{figure}

 The total number of automata in this class is $(2 \times 2)^{2 \times 2}=256$: [2,2,0,0], [2,2,0,1], $\ldots$, [2,2,15,15]. In this class, the number of bijective ones is given by $4!=24$. Among these, 
 \begin{itemize}
  \item the number of transitive ones is 20, 
  \item the number of particle-preserving ones is 6, 
  \item and the number of transitive and particle-preserving ones is 5: MA-[2,2,5,3], [2,2,6,5], [2,2,9,5], [2,2,10,12], and [2,2,12,5].
 \end{itemize}
 The automata [2,2,6,5], [2,2,9,5], and [2,2,12,5] identically act on binary sequences, so these are not minimal. The automata [2,2,5,3] and [2,2,10,12] are isomorphic by exchanging two states $q_0$ and $q_1$. Therefore, [2,2,5,3] ([2,2,10,12]) is the only automaton that is bijective, transitive, particle-preserving, and minimal. This automaton is known as the box--ball system with a carrier capacity of one (BBS-C(1)). Figure~\ref{Fig: BBS-C(1)} shows the state translation diagram of MA-[2,2,5,3]. 

 \subsection{\texorpdfstring{Case $Q=\{q_0,q_1,q_2\}, S=\{0,1\}$}{}}
 The total number of automata in this class is $(3 \times 2)^{3 \times 2}=46656$: [3,2,0,0], [3,2,0,1], $\ldots$, [3,2,728,63]. In this class, the number of bijective ones is given by $6!=720$. Among these, 
 \begin{itemize}
  \item the number of transitive ones is 592, 
  \item the number of particle-preserving ones is 84, 
  \item and the number of transitive and particle-preserving ones is 68.
 \end{itemize}
 42 out of these 68 automata are minimal, and they can be divided into seven types up to the permutation of three states: [3,2,104,11], [3,2,146,7], [3,2,128,19], [3,2,154,7], [3,2,318,52], [3,2,106,13], [3,2,266,19]. Assuming the vacuum alphabet $V=\{0\}$, we found only three types of locally interacting automata:
 \begin{enumerate}
  \renewcommand{\labelenumi}{(\arabic{enumi})}
  \item MA-[3,2,104,11]\quad (BBS-C(2), carrier capacity two), 
  \item MA-[3,2,146,7]\quad (BBS-V(2), jump on to the secondary nearest vacant box), 
  \item MA-[3,2,154,7]\quad (BBS-S(2), skip to the box two spaces ahead).
 \end{enumerate}
 These three automata are candidates for soliton automata. Figure~\ref{Fig4: MA-type(3,2)} shows the state translation diagrams of these automata. The BBS-C(2) is known to be a soliton automaton as a box--ball system with a carrier capacity of two. We will discuss the integrability of the other candidates in the following section.
 \begin{figure}[ht]\centering
  \includegraphics[width=0.63\linewidth]{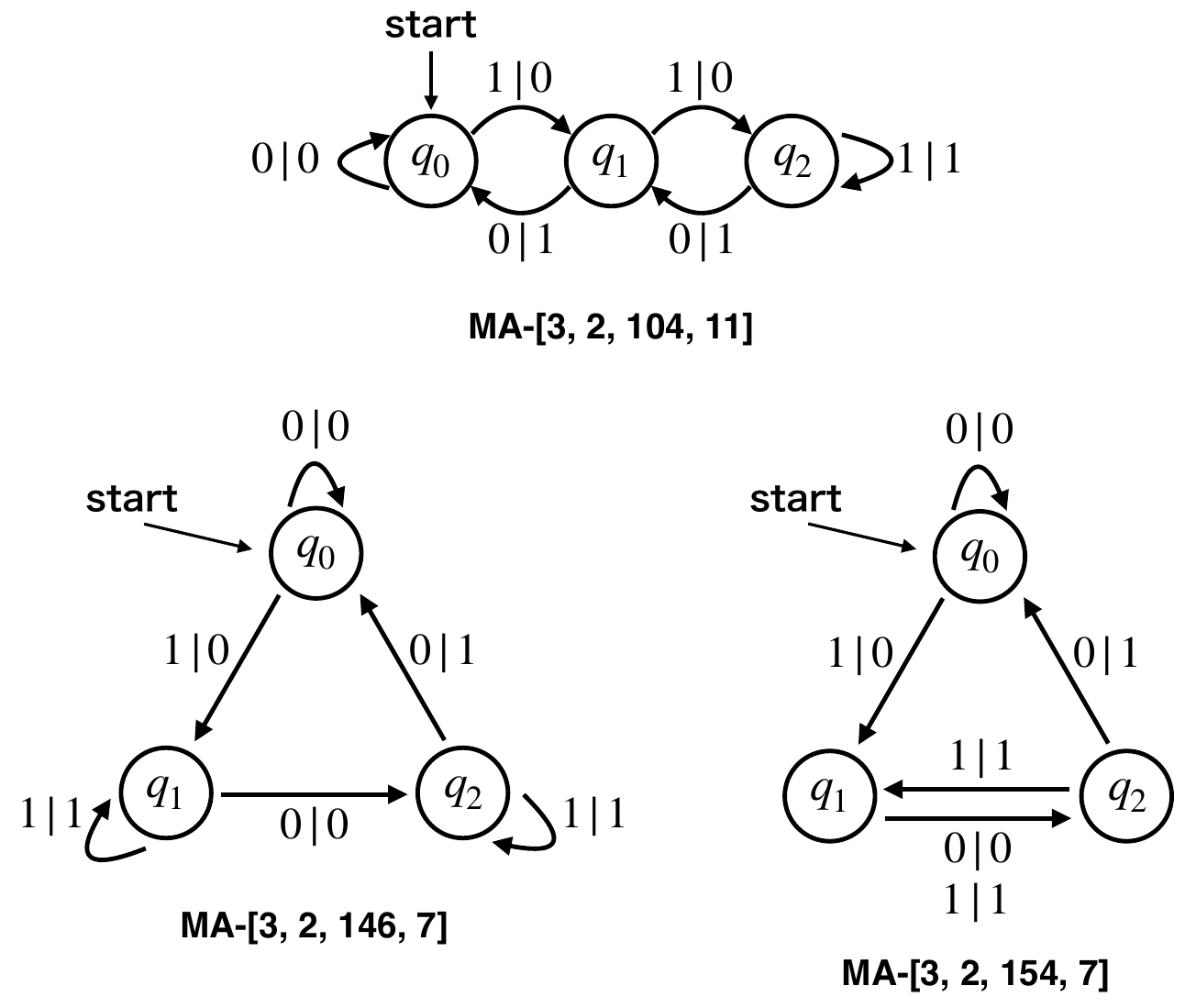}
  \caption{The bijective, transitive, particle-preserving, minimal, and locally interacting Mealy automaton of type (3, 2). }
  \label{Fig4: MA-type(3,2)}
 \end{figure}

 \section{Linearization of the MA-[3,2,154,7] (BBS-S(2))}
 In this section, we interpret the MA-[3,2,154,7] as the box--ball system and prove the integrability by introducing the linearization method. 

 \subsection{\texorpdfstring{BBS-S($2$): skip to the box two spaces ahead}{}}
 Let us introduce the time evolution of the BBS-S(2): the BBS of skipping to the box two spaces ahead rule. Consider the boxes aligned in a semi-infinite row and finitely many balls in some boxes. The time evolution of the BBS-S($2$) is described using a carrier as follows: 
 \begin{enumerate} \renewcommand{\labelenumi}{(\roman{enumi})}
  \item The carrier of capacity one moves from the left end to the right. 
  \item The empty carrier moves right by one box until meeting the box occupied by a ball. If the carrier is empty and there are no balls in the boxes right to the carrier, then the carrier stops. 
  \item When the empty carrier meets the box occupied by a ball, take the ball from the box to the carrier and move to the right by skipping every one box until meeting the vacant box. Meeting the vacant box, the carrier unloads the ball to the vacant box and follows step (ii). 
 \end{enumerate}

 The MA-[3,2,154,7] with the initial state $q=q_0$ corresponds to the BBS-S(2). The automaton is given by
 \begin{equation}
  [Q,S,\phi,\psi]= \left[\{q_0, q_1, q_2\},\{0, 1\},\left( \begin{array}{ccc}0 &1\\ 2 &2\\ 0 &1\end{array}\right), \,\left( \begin{array}{ccc}0&0\\ 0&1\\ 1&1 \end{array}\right)\right].
 \end{equation}

 \subsection{Bijection}
 We construct a bijection between the BBS-S(2) state sequence and two sequences of nonnegative integers. 
 \begin{pro}
  Let $S=\{0,1\}$.
  \begin{equation}
   S^* \cong \{(b_1,b_2,\ldots,b_n) \mid n\in\Z_{\ge0}, b_j \in \Z_{\ge0}, b_1\le b_2\le\cdots\le b_n\}\times \{(c_1,c_2,\ldots,c_n) \mid n\in\Z_{\ge0}, c_j \in \Z_{\ge0}, c_1<c_2<\cdots<c_n\}.
  \end{equation}
 \end{pro}
 The procedure for computing two sequences from $\Bs\in S^*$ is as follows: 
 \begin{enumerate}
  \item For $\Bs\in S^*$, replace contiguous `1's with `b' or `c' as follows:
  \begin{eqnarray}
\begin{cases}
   1^{2k}\mapsto \rmb^k &(\mbox{the length of `1's is even}),\\
   1^{2k+1}\mapsto \rmb^k\rmc &(\mbox{the length of `1's is odd}).
\end{cases}  
  \end{eqnarray}
  Here, when the length of `1's is odd, the position of `c' is not necessarily at the end of `b's, that is, $1^{2k+1}$ can be replaced with $\rmb^j\rmc\rmb^{k-j}$ instead of $\rmb^k\rmc$. For example, 
  \begin{eqnarray}
   1000100100011110011011011110000\cdots&\mapsto& \rmc000\rmc00\rmc000\rmb\rmb00\rmb0\rmb0\rmb\rmb0000\cdots\nonumber\\
   0000001011110011101101111100000\cdots&\mapsto& 000000\rmc0\rmb\rmb00\rmc\rmb0\rmb0\rmc\rmb\rmb00000\cdots\nonumber
  \end{eqnarray}
  \item We define two sequences $\overline{b}=(b_1, b_2, \ldots)$ and $\overline{c}=(c_1, c_2, \ldots)$ from the sequence of $\{0, \rmb, \rmc\}$ obtained above. Let $b_j$ denote the number of 0s on the left-side hand of the $j$th $b$ from the left. Similarly, let $c_j$ denote the number of 0s on the left-hand side of the $j$th $c$ from the left. 
  \begin{eqnarray}
   1000100100011110011011011110000\cdots&\mapsto& \rmc000\rmc00\rmc000\rmb\rmb00\rmb0\rmb0\rmb\rmb0000\cdots\nonumber\\
   &\mapsto& 
\begin{cases}
 \overline{b}=(8, 8, 10, 11, 12, 12),\\ \overline{c}=(0, 3, 5). 
\end{cases}
\nonumber\\
   0000001011110011101101111100000\cdots&\mapsto& 000000\rmc0\rmb\rmb00\rmc\rmb0\rmb0\rmb\rmb\rmc00000\cdots \nonumber\\&\mapsto& 
\begin{cases}
\overline{b}=(7, 7, 9, 10, 11, 11),\\ \overline{c}=(6, 9, 11). 
\end{cases}
\nonumber
  \end{eqnarray}
  By definition, $\overline{b}$ is an increasing integer sequence, and $\overline{c}$ is a strictly increasing integer sequence. 
 \end{enumerate}

 \subsection{Linearization}
 There are two types of solitary waves as
 \begin{itemize}
  \item The solitary wave of speed one: $0\rmb^k0\mapsto 00\rmb^k \quad(0(11)^k0 \mapsto 00(11)^k)$,
  \item The solitary wave of speed two: $0(\rmc0)^k0\mapsto 00(0\rmc)^k \quad(0100 \mapsto 0001)$. 
 \end{itemize}

 We slightly modify the rule for the time evolution of the BBS-S(2). Here, when considering the $\{\rmb, \rmc\}$ sequence corresponding to $1^{2k+1}$, `$\rmc$' is assumed to be the rightmost. After the carrier loads a ball from the box, it unloads the ball into the second box. If the second box contains another ball, it is exchanged for the loaded ball. After that, the balls associated with each `$\rmc$' are exchanged with the balls associated with `$\rmb$' to the right of `$\rmc$' until the number of letters to the right of `$\rmc$' becomes zero. This change in procedure does not affect the final $\{0, 1\}$ sequence, as it only replaces `1' with another `1'. Under this time evolution rule, each `1' either moves to the second box or is overtaken by other balls. Therefore, the following holds. 
 \begin{itemize}
  \item Every `1' in `c' moves to the second box. Because the character to the right of `c' is `0', it does not overtake other `1's. Since the number of `1's on the left side of it does not change, the number of `0's on the left side increases by two. 
  \item The first `1' in each `b' overtakes the second `1' and moves to the second box. Since the number of `1's on the left side of it increases by one, the number of `0's on the left side increases by one.
  \item The second `1' in each `b' is overtaken by the first `1' and does not move. The number of `1's on the left side of it decreases by one.
 \end{itemize}
 The first `1's in `b' becomes the second `1's in `b' the next time, and the second `1's becomes the first `1's. Thus, for every `b', the number of balls on the left side increases by one. Let $\overline{b}^{t}=(b_1^t, b_2^t,\ldots)$ and $\overline{c}^t=(c_1^t, c_2^t, \ldots)$ be the corresponding integer sequences at time $t$. Then, we find 
 \begin{eqnarray}
  b_j^t &=& b_j^0 + t,\\
  c_j^t &=& c_j^0 + 2t.
 \end{eqnarray}

 \begin{exm}
  In the following series of the time evolution of the BBS-S(2), a solitary wave with speed one is overtaking a solitary wave with speed two. We can observe that the behavior of the `10' (`11') sequence becomes a simple parallel translation from left to right after the interaction. 

\begin{align*}
 \begin{array}{llll}
&t=\phantom{1}0:101001001111001101100000000000000000000, &\overline{b}^{0}=(5,5,7,8),&\overline{c}^{0}=(0,1,3)\\
&t=\phantom{1}1:001010010111100110110000000000000000000, &\overline{b}^{1}=(6, 6, 8, 9),&\overline{c}^{1}=(2, 3, 5)\\
&t=\phantom{1}2:000010100111110011011000000000000000000, &\overline{b}^{2}=(7, 7, 9, 10),&\overline{c}^{2}=(4, 5, 7)\\
&t=\phantom{1}3:000000101011110101101100000000000000000, &\overline{b}^{3}=(8, 8, 10, 11),&\overline{c}^{3}=(6, 7, 9)\\
&t=\phantom{1}4:000000001011111001110110000000000000000, &\overline{b}^{4}=(9, 9, 11, 12),&\overline{c}^{4}=(8, 9, 11)\\
&t=\phantom{1}5:000000000011111010110111000000000000000, &\overline{b}^{5}=(10, 10, 12, 13),&\overline{c}^{5}=(10, 11, 13)\\
&t=\phantom{1}6:000000000001111010111011010000000000000, &\overline{b}^{6}=(11, 11, 13, 14),&\overline{c}^{6}=(12, 13, 15)\\
&t=\phantom{1}7:000000000000111100111011100100000000000, &\overline{b}^{7}=(12, 12, 14, 15),&\overline{c}^{7}=(14, 15, 17)\\
&t=\phantom{1}8:000000000000011110011011101001000000000, &\overline{b}^{8}=(13, 13, 15, 16),&\overline{c}^{8}=(16, 17, 19)\\
&t=\phantom{1}9:000000000000001111001101101010010000000, &\overline{b}^{9}=(14, 14, 16, 17),&\overline{c}^{9}=(18, 19, 21)\\
&t=10:000000000000000111100110110010100100000, &\overline{b}^{10}=(15,15,17,18),&\overline{c}^{10}=(20,21,23)\\
&t=11:000000000000000011110011011000101001000, &\overline{b}^{11}=(16,16,18,19),&\overline{c}^{11}=(22,23,25)\\
\end{array}  
\end{align*}
 \end{exm}

 \section{Integrability of MA-[3,2,146,7] (BBS-V(2))}
 In this section, we interpret MA-[3,2,146,7] as the box--ball system, and prove the time evolution of the BBS-V(2) is equivalent to the ultradiscrete Lotka--Volterra (uLV) equation. 
 
 \subsection{\texorpdfstring{BBS-V(2): jump on to the second nearest vacant box rule}{}}
 Let us introduce the time evolution of the BBS-V(2): the BBS of jumping to the second nearest vacant box rule. Consider the boxes aligned in a semi-infinite row and finitely many balls in some boxes. The time evolution of BBS-V($2$) is described using a carrier as follows: 
 \begin{enumerate} \renewcommand{\labelenumi}{(\roman{enumi})}
  \item The carrier of capacity one moves from the left end to the right. 
  \item The empty carrier moves right by one box until meeting the box occupied by a ball.  
  \item When the empty carrier meets the box occupied by a ball, the carrier takes the ball from the box and jumps onto the second right nearest vacant box. Then the carrier unloads the ball to the vacant box and follows step (ii). 
 \end{enumerate}
 The automaton MA-[3,2,146,7] with the initial state $q=q_0$ corresponds to the BBS-V(2). The automaton is given by
 \begin{equation}
  [Q,S,\phi,\psi]= \left[\{q_0, q_1, q_2\},\{0, 1\},\left( \begin{array}{ccc}0 &1\\ 2 &1\\ 0 &2\end{array}\right), \,\left( \begin{array}{ccc}0&0\\ 0&1\\ 1&1 \end{array}\right)\right].
 \end{equation}

 \begin{exm}
  In the following time evolution series of the BBS-V(2), there are three solitons consisting of one, two, and five balls, respectively. A soliton of $n$ balls moves by $n+1$ boxes in $n$ time steps. Therefore, the speed of a soliton consisting of $n$ balls is $(n+1)/n$. 
\begin{align*}
 \begin{array}{llll}
&t=\phantom{1}0:1001100000111110000000000000000000\\
&t=\phantom{1}1:0010101000011110100000000000000000\\
&t=\phantom{1}2:0000110010001110110000000000000000\\
&t=\phantom{1}3:0000010100100110111000000000000000\\
&t=\phantom{1}4:0000000110001010111100000000000000\\
&t=\phantom{1}5:0000000010100011011101000000000000\\
&t=\phantom{1}6:0000000000110001011110010000000000\\
&t=\phantom{1}7:0000000000010100011111000100000000\\
&t=\phantom{1}8:0000000000000110001111010001000000\\
&t=\phantom{1}9:0000000000000010100111011000010000\\
&t=10:0000000000000000110011011100000100\\
&t=11:0000000000000000010101011110000001\\
\end{array}  
\end{align*}
 \end{exm}

 \subsection{The BBS-V(2) and the original BBS}
  One of the authors proved that the time evolution of a shifted BBS-V(2) is connected to the time evolution of the original BBS via the ultradiscrete Lotka--Volterra (uLV) equation~\cite{Y06}. First, we recall the relationships between the original BBS and the uLV equation~\cite{TH98}. The time evolution of the original BBS introduced in 1990~\cite{TS90} is described by the piecewise linear equation called the ultradiscrete Korteweg--de Vries (uKdV) equation:
 \begin{equation}
  \eta_n^{t+1} = \min\left\{1-\eta_n^t, \sum_{i=-\infty}^{n-1}(\eta_i^t-\eta_i^{t+1})\right\}.
 \end{equation}
 Through the variable transformation
 \begin{equation}
  \gamma_n^t = \sum_{j=-\infty}^{n+1}\eta_j^t - \sum_{j=-\infty}^n\eta_j^{t+1},
 \end{equation}
 we have the uLV equation
 \begin{equation}
  \gamma_{n+1}^{t+1}-\gamma_n^t=\max(0, \gamma_n^{t+1}-1)-\max(0, \gamma_{n+1}^t-1). 
  \label{eq:uLV}
 \end{equation}

 Next, we explain the correspondence between the BBS-V(2) and the uLV equation in~\cite{Y06}. There are two types of input/output sequences with state transitions from $q_0$ to $q_0$:
 \begin{enumerate}
  \item input ``0'', output ``0'',
  \item input ``$1^{n+1}01^m0$'', output ``$01^n01^{m+1}$'',
 \end{enumerate}
 where $n, m\geq0$, and $1^n$ denotes the sequence $\underbrace{11\cdots1}_{n}$.

 \begin{pro}
  Let $X=\{0\}\cup\{1^{n+1}01^m0\}_{n, m\geq0}$ be the set of sequence. Then, any sequence $s00\,(s\in S^*)$, which has suffix $00$, corresponds one-to-one with an element of $X^*$. 
 \end{pro}
 Because we assume that the input sequence has finitely many 1s and there are sufficiently many 0s at the right end of the sequence, the sequence can be divided uniquely as $x_1x_2\cdots x_n\,(x_i\in X)$. Let $Y=\{0\}\cup\{(n+1, m)\}_{n, m\geq0}\subset\Z\cup\Z^2$, and define a map $\mu:X\to Y$ as $0\mapsto0(\in\Z)$ and $1^{n+1}01^m0\mapsto(n+1, m)(\in\Z^2)$. Introduce the extension of $\mu$ on $X^*$ by
 \begin{equation}
  \mu(x_1x_2\cdots x_n)=\mu(x_1)\mu(x_2)\cdots\mu(x_n).
  \label{eq: mu}
 \end{equation}
 The product of sequences of integers in the right-hand side of the equation (\ref{eq: mu}) denotes the concatenation of sequences. 

 \begin{exm}
  The sequence 
  \begin{equation*}
   0\,0\,1\,1\,1\,0\,1\,0\,1\,0\,0\,0\,1\,1\,0\,1\,1\,0\,0\cdots\in S^*
  \end{equation*}
  is divided as 
  \begin{equation*}
   \underline{0}\,\underline{0}\,\underline{1\,1\,1\,0\,1\,0}\,\underline{1\,0\,0}\,\underline{0}\,\underline{1\,1\,0\,1\,1\,0}\,\underline{0}\cdots\in X^*.
  \end{equation*}
  Applying $\mu$ for every underlined subsequence gives us 
  \begin{equation*}
   \underline{0}\,\underline{0}\,\underline{3\,1}\,\underline{1\,0}\,\underline{0}\,\underline{2\,2}\,\underline{0}\cdots\in Y^*,
  \end{equation*}
  and we finally get the sequence 
  \begin{equation*}
   0\,0\,3\,1\,1\,0\,0\,2\,2\,0\cdots\in\Z^*.
  \end{equation*}
  \label{Ex: mu}
 \end{exm}

\begin{figure}[htbp]
 \centering
 \includegraphics[height=1.8cm]{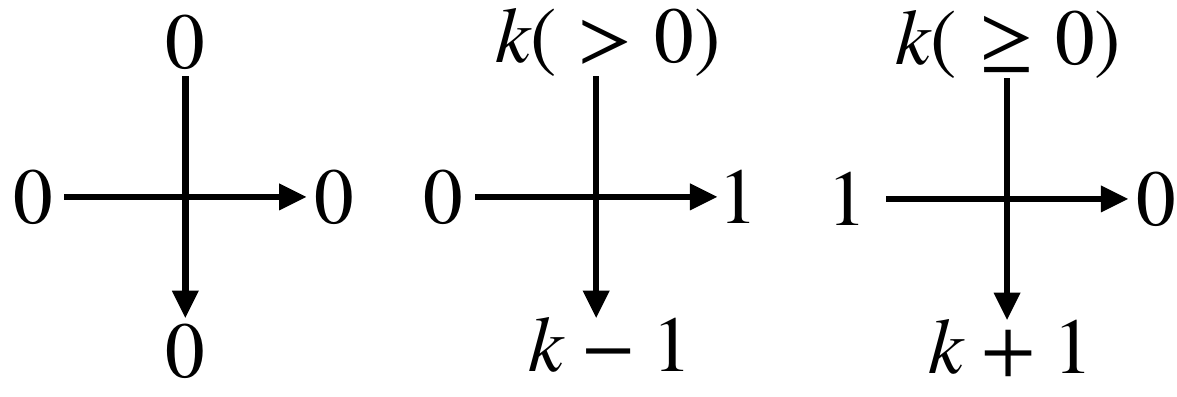}
 \caption{The action of $T$ on $\{0, 1\}\times\Z_{\geq0}$. }
 \label{fig:op_T}
\end{figure}

 Introduce the operator $T$ on lattice (in Figure \ref{fig:op_T}) as
 \begin{eqnarray}
  T&:& \{0, 1\}\times\Z_{\geq0}\to\Z_{\geq0}\times\{0, 1\}, \nonumber\\
  &:& (i, j)\mapsto (j^\prime, i^\prime)=(\max(2i+j-1, 0), \min(1-i, j)) .
  \label{eq:def_T}
 \end{eqnarray}

\begin{figure}[htbp]
 \centering
 \includegraphics[height=1.8cm]{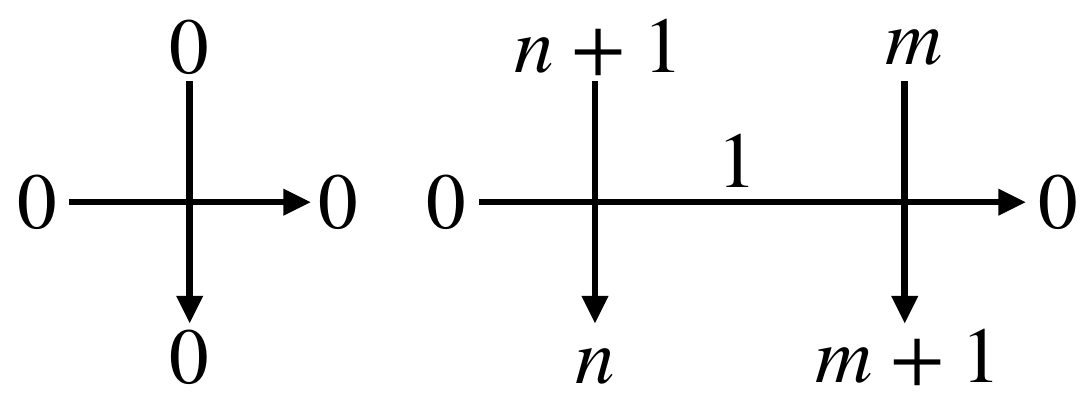}
 \caption{The action of $T$ and $(T\otimes {\bm 1})({\bm 1}\otimes T)$ that correspond to MA-[3,2,146,7].}
 \label{fig:action_T}
\end{figure}

\noindent
 The operator $T$ satisfies
 \begin{eqnarray*}
  T&:& (0, 0)\mapsto(0, 0),\\
  (T\otimes {\bm 1})({\bm 1}\otimes T)&:&(0, n+1, m)\mapsto(n, m+1, 0),
 \end{eqnarray*}
 and these correspond to the input/output of the MA-[3,2,146,7] by shifting one cell to the left, where ${\bm 1}:\Z_{\geq0}\to\Z_{\geq0}$ is the identity operator. Both two operators $T$ and $(T\otimes{\bm 1})({\bm 1}\otimes T)$ can be regarded as the map $\{0, 1\}\times Y\to Y\times\{0, 1\}$ shown in Figure \ref{fig:action_T}. 

 Define the operator $\mcT$ as 
 \begin{eqnarray}
  \mcT&=& \mcT^{(n-1)}\cdots\mcT^{(2)}\mcT^{(1)}\mcT^{(0)},\\
  \mcT^{(i)} &=& \underbrace{{\bm 1}\otimes\cdots\otimes{\bm 1}}_{i}\otimes\,T\!\otimes\underbrace{{\bm 1}\otimes\cdots\otimes{\bm 1}}_{n-i-1}.
 \end{eqnarray}

 \begin{pro}
  Let $\Delta$ be the map $S^*\to S^*$ of the time evolution of the MA-[3,2,146,7]. Then, $\mcT\circ\mu\simeq\mu\circ\Delta$ with shifting one cell to the left. 
 \end{pro}

\begin{figure}[htbp] \centering
  \includegraphics[height=1.8cm]{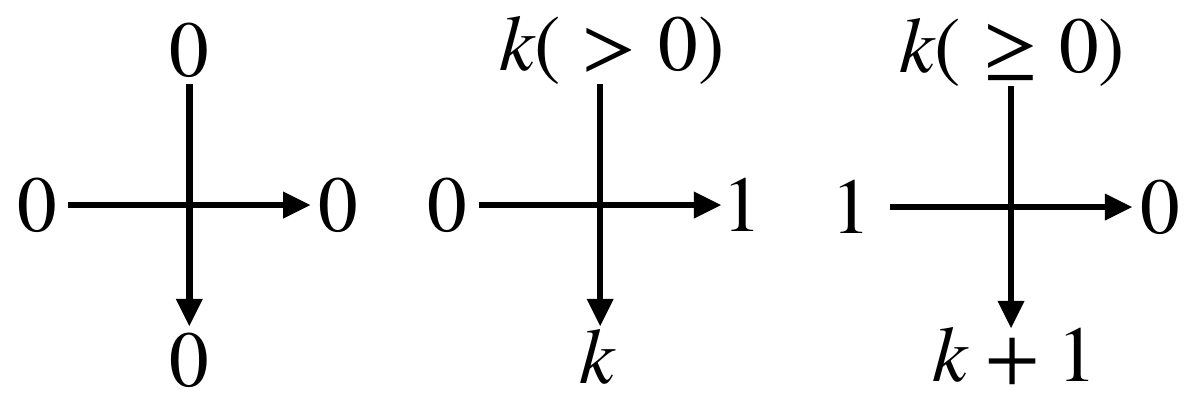}
  \caption{The action of $U$ on $\{0, 1\}\times\Z_{\geq0}^*$.}
  \label{fig:op_S}
 \end{figure}

 Next, introduce an operator $U$ (in Figure \ref{fig:op_S}) like $T$ as
 \begin{eqnarray}
  U&:& \{0, 1\}\times\Z_{\geq0}\to\Z_{\geq0}\times\{0, 1\},\nonumber\\
  &:& (i, j)\mapsto (j^\prime, i^\prime)=(i+j, \min(1-i, j)).
  \label{eq:def_S}
 \end{eqnarray}

\begin{figure}[htbp] \centering
  \includegraphics[height=1.8cm]{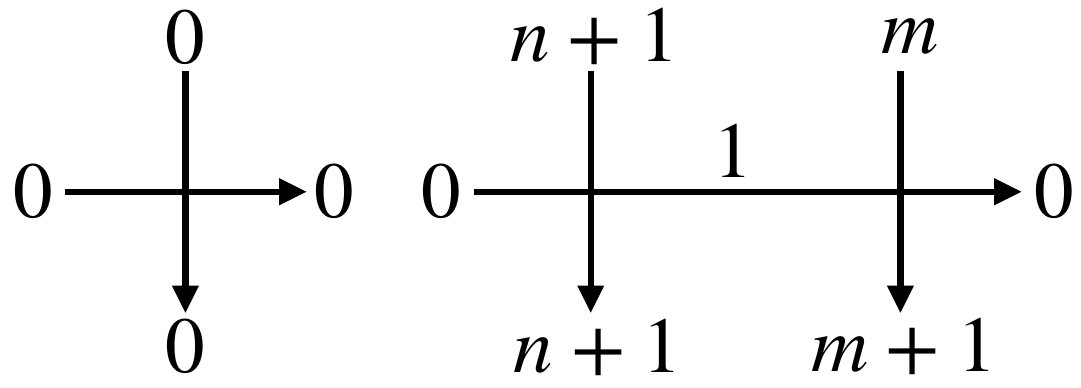}
  \caption{The action of $U$ and $(U\otimes{\bm 1})({\bm 1}\otimes U)$.}
  \label{fig:action_S}
 \end{figure}

\noindent
 The operator $U$ satisfies
 \begin{eqnarray*}
  U&:& (0, 0)\mapsto(0, 0),\\
  (U\otimes{\bm 1})({\bm 1}\otimes U)&:&(0, n+1, m)\mapsto(n+1, m+1, 0),
 \end{eqnarray*}
 and both $U$ and $(U\otimes {\bm 1})({\bm 1}\otimes U)$ can be regarded as the map $\{0, 1\}\times Y\to Y\times\{0, 1\}$ shown in Figure \ref{fig:action_S}.

 Define the operator $\mcU$ like $\mcT$ as
 \begin{eqnarray}
  \mcU &=& \mcU^{(n-1)}\cdots\mcU^{(2)}\mcU^{(1)}\mcU^{(0)},\\
  \mcU^{(i)} &=& \underbrace{{\bm 1}\otimes\cdots\otimes {\bm 1}}_{i}\otimes\,U\!\otimes\underbrace{{\bm 1}\otimes\cdots\otimes{\bm 1}}_{n-i-1}.
 \end{eqnarray}
 Then, the following theorem holds.
 \begin{thm}
  Let ${\rm uLV}$ be the map $Y^*\to Y^*$ of the time evolution of the uLV equation given by the equation (\ref{eq:uLV}). Then, ${\rm uLV}\circ\mcU\simeq\mcU\circ\mcT$. 
  \begin{proof}
   For $x\in\Z_{\geq0}^*$, let $y=\mcU x$, $x^\prime=\mcT x$, and $y^\prime=\mcU x^\prime\in\Z_{\geq0}^*$. We will prove that the map $y\to y^\prime$ satisfies the uLV equation (\ref{eq:uLV}).

   The terms of $x$ are denoted by $x=(x_1, x_2, x_3, \ldots), x_i\in\Z_{\geq0}^*$, and the same applied to $y$, $x^\prime$, and $y^\prime$. In addition, define the auxiliary variables for $y=\mcU x$, $x^\prime=\mcT x$ and $y=\mcU x^\prime$ as in Figure \ref{fig:ThuLV}. Note that the auxiliary variables for $\mcU$ are the same as those of $\mcT$ (see Figure \ref{fig:op_T} and Figure \ref{fig:op_S}). 
   \begin{figure}[htbp] \centering
    \includegraphics[height=6cm]{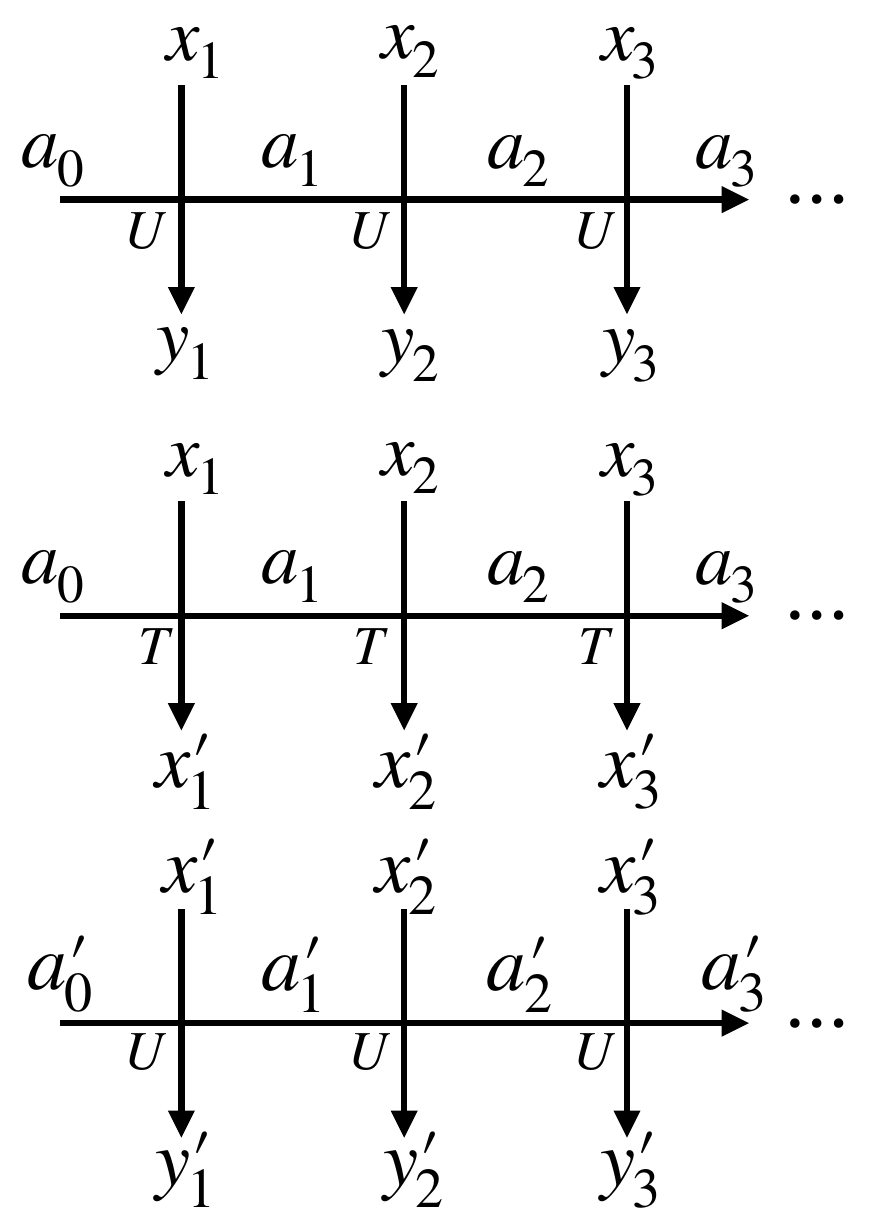}
    \caption{The auxiliary variables $a_i, a_i^\prime$.}
    \label{fig:ThuLV}
   \end{figure}
   We can set $a_0=a_0^\prime=0$ for the boundary condition.

   \noindent Now, we have 
   \begin{eqnarray}
    y_{i+1} &=& x_{i+1}+a_i,\label{eq:Sx}\\
    y^\prime_{i+1} &=& x^\prime_{i+1}+a^\prime_i\label{eq:Sx^p}
   \end{eqnarray}
   from (\ref{eq:def_S}) (the definition of $U$), and 
   \begin{equation}
    x^\prime_{i+1} = x_{i+1}+a_i-a_{i+1}\label{eq:Tx}
   \end{equation}
   from (\ref{eq:def_T}) (the definition of $T$) and $a_i, x_i\in\Z_{\geq0}$. Here, by using $a_j\in\{0, 1\}$, we obtain 
   \begin{eqnarray}
    \min(y_i, 1) &=& \min(x_i+a_{i-1}, 1)\nonumber\\
    &=& a_{i-1}+\min(x_i, 1-a_{i-1})\nonumber\\
    &=& a_{i-1}+a_i, \label{eq:con1}
   \end{eqnarray}
   and
   \begin{equation}
    \min(y^\prime_i, 1)=a^\prime_{i-1}+a^\prime_i\label{eq:con2}
   \end{equation}
   in the same way. The equations (\ref{eq:Sx})--(\ref{eq:con2}) lead us to 
   \begin{eqnarray}
    y^\prime_{i+1}+\max(y^\prime_i-1, 0) &=& y^\prime_{i+1}+y^\prime_i-\min(1, y^\prime_i)\nonumber\\
     &=& y^\prime_{i+1}+y^\prime_i-a^\prime_{i-1}-a^\prime_i\nonumber\\
     &=& x^\prime_{i+1}+x^\prime_i\nonumber\\
     &=& (x_{i+1}+a_i-a_{i+1})+(x_i+a_{i-1}-a_i)\nonumber\\
     &=& y_{i+1}+y_i-a_{i+1}-a_i\nonumber\\
     &=& y_{i+1}+y_i-\min(y_{i+1}, 1)\nonumber\\
     &=& y_i+\max(y_{i+1}-1, 0).
   \end{eqnarray}
   Setting $y_i^t=y_i, y_i^{t+1}=y^\prime_i, \gamma^i_t=y^t_i$, the variables $\gamma^t_i$ satisfies the uLV equation (\ref{eq:uLV}). 
  \end{proof}
  \begin{cor}
   ${\rm uLV}\circ(\mcU\circ\mu)\simeq(\mcU\circ\mu)\circ\Delta$. 
  \end{cor}
 \end{thm}

 \section{Concluding Remarks}
 In this paper, focusing on the BBS with finite carrier capacity, we introduced several key properties of Mealy automata, including particle-preserving, bijective, transitive, and locally interacting. Through computational and theoretical analysis based on these properties, we identified three classes of 3-state soliton Mealy automata over a 2-letter alphabet: MA-[3,2,104,11], MA-[3,2,146,7], and MA-[3,2,154,7]. 
 
 MA-[3,2,104,11] corresponds to the BBS with a carrier capacity of two (BBS-C(2)). The BBS-C(2) has been extensively studied and its time evolution can be linearized~\cite{KNTW18, KOSTY06}. MA-[3,2,146,7] (BBS-V(2)) is described as the secondary nearest vacant box rule, while MA-[3,2,154,7] (BBS-S(2)) is described as the skipping to the box two spaces ahead rule. For the BBS-S(2), we provided a simple method for linearizing its time evolution. Furthermore, we showed that the time evolution of the BBS-V(2) is equivalent to the ultradiscrete Lotka--Volterra equation, which is closely related to Takahashi--Satsuma's box--ball system. 
 
 For Mealy automata with a state set size of $|Q|\ge4$, candidates for soliton automata that satisfy the above key properties can be enumerated using a computer. By computing the time evolution starting from several initial binary sequences, we checked whether they are soliton automata. It was confirmed that some candidates do not exhibit solitonic properties. A theoretical analysis of these candidates also remains for future work. 

 \section*{Acknowledgments}
 The research of ST is supported by JSPS KAKENHI (Grant Number JP24K00528) and
the research of FY is supported by JSPS KAKENHI (Grant Number JP23K03233).  
This research is partially supported by the joint project ``Advanced Mathematical Science for Mobility Society'' of Kyoto University and Toyota Motor Corporation.

 
\end{document}